# The finite tiling problem is undecidable in the hyperbolic plane

Maurice Margenstern[1]

Laboratoire d'Informatique Théorique et Appliquée, EA 3097,
Université de Metz, I.U.T. de Metz,
Département d'Informatique,
Île du Saulcy,
57045 Metz Cedex, France,
`margens@univ-metz.fr`

**Abstract.** In this paper, we consider the finite tiling problem which was proved undecidable in the Euclidean plane by Jarkko Kari, see [4]. Here, we prove that the same problem for the hyperbolic plane is also undecidable.

## 1 Introduction

A lot of problems deal with tilings. Most of them are considered in the setting of the Euclidean plane. A certain number of these problems turn out to be undecidable in this frame, thanks to the facility to simulate the computation of a Turing machine in this setting. The most famous case of such a problem is the **general tiling problem** proved to be undecidable by Berger in 1964, see [1]. In 1971, R. Robinson gave a simplified proof of the same result, see [16]. Sometimes, the general problem is simply called the **tiling problem**. The reason of these different names lies in the fact that several conditions were put on the problem, leading to different settings, and a dedicated proof was required each time when the problem turned out to be undecidable. Among these variations, the most well-known is the **origin-constrained** problem, proved to be undecidable by Wang in 1958, see [18].

The general tiling prolem consists in the following. Given a finite set of tiles $T$, is there an algorithm which says whether it is possible to tile the plane with copies of the tiles of $T$ or not? The **origin-constrained** problem consists in the same question to which a condition is appended: given a finite set of tiles $T$ and a tile $T_0 \in T$, is there an algorithm which says whether i is possible or not to tile the plane with copies of the tiles of $T$ or not, the first tile being $T_0$? In the general problem there is no condition on the first tile: it can be a copy of any tile of $T$.

There are a lot of variants of these problems and the reader is referred to [16], where an account is given on several such conditions.

The **finite tiling problem** is a slightly different question. Given a finite set of tiles $T$ and a tile $b$ with $b \notin T$, called the **blank**, is there an algorithm which says whether there is a tiling of the plane with copies of $T \cup \{b\}$ in which there are only finitely many copies of tiles of $T$ but at least one? This problem was first formulated by J. Kari in [4], where it is proved to be undecidable. In the case of the general tiling problem, a **solution** to the problem is a tiling of the plane with copies of the tiles from $T$ only. In the case of the finite tiling problem, a **solution** to the problem is a tiling in which only finitely many copies of $T$ are used and at least one is used.

The general tiling problem for the hyperbolic plane was raised by R. Robinson in his 1971 paper, see [16]. In 1978, R. Robinson proved that the origin-constrained problem is undecidable in the hyperbolic plane, see [17]. The general tiling problem for the hyperbolic plane remained pending a long time. In 2006, the present author proved that the tiling problem with an intermediate condition, so called **generalized origin-constrained problem** is undecidable, see [8,9]. Recently, the present author proved the general tiling problem to be undecidable in the hyperbolic plane, see [10,13]. At the same time, J. Kari established the same result, using a completely different approach, see [5].

In this paper, we prove that:

**Theorem 1** *The finite tiling problem is undecidable in the hyperbolic plane.*

As we shall see, the solution makes use of the technique used in [8,9].

In the next section, section 2, we sketchilly remind the solution to the origin-constrained problem which we have given in [8,9]. In section 3, we prove the theorem. In the conclusion, we discuss a few possible issues.

## 2  The harp

The solution of the origin-constrained problem which we now consider takes place in the tiling $\{7,3\}$ of the hyperbolic plane. It consists in simulating the space-time diagram of a Turing machine.

Now, we turn to our first sub-section, where we remind what we need of the tiling $\{7,3\}$ of the hyperbolic plane.

### 2.1  The tiling $\{7,3\}$ of the hyperbolic plane

The tiling $\{7,3\}$ is obtained from the regular heptagon with an interior angle of $\dfrac{2\pi}{3}$ by reflection in its edges and, recursively, by reflection of the images in their edges. The existence of the tiling is a corollary of Poincaré's theorem on a sufficient conditions for tiling the hyperbolic plane by triangles. It is enough to consider the rectangluar triangle of the hyperbolic plane with the acute angles $\dfrac{\pi}{7}$ and $\dfrac{\pi}{3}$. Below, figure 2 illustrates the tiling $\{7,3\}$.

In [2], we introduced a way to exhibit a generating tree of the tiling which is basically the same as the generating tree of the pentagrid, the tiling $\{5,4\}$ of the hyperbolic plane. This tiling is constructed by a process, similar to the one used for constructing the tiling {7,3}, but it is used with the regular rectangluar pentagon. This tree is called the **standard Fibonacci tree**, simply **Fibonacci tree** in the sequel, see [6] for more details on this tree.

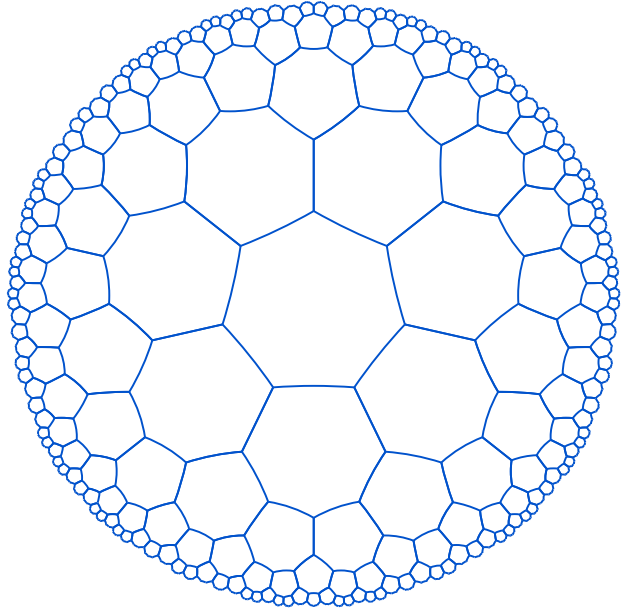

**Figure 1** *The tiling {7,3} of the hyperbolic plane in the Poincaré's disc model.*

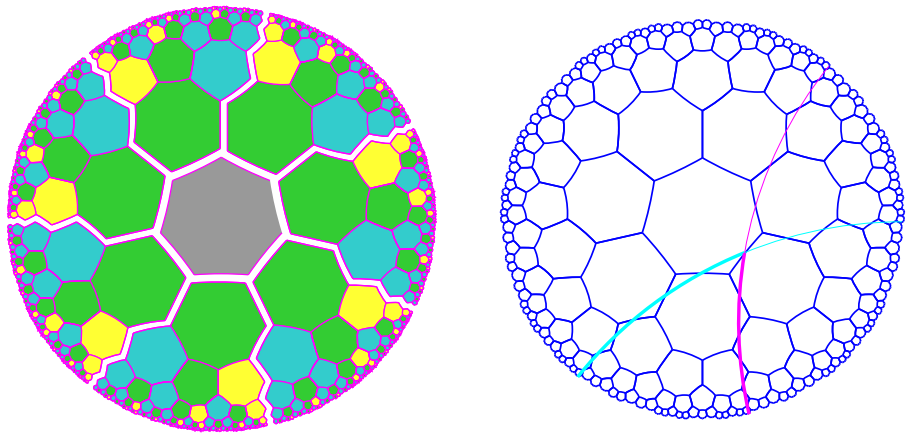

**Figure 2** *Left-hand side: the standard Fibonacci trees which span the tiling {7,3} of the hyperbolic plane. Right-hand side: the mid-point lines.*

The way to exhibit the Fibonacci tree is based on the **mid-point lines**, which we introduced in [2]. As suggested by their name, these lines join the mid-points of two consecutive edges of a heptagon. It turns out that the angular sector determined by two rays obtained by two mid-point lines meeting at a mid-point $C$, joining the two mid-points of the two other edges which meet at $C$, exactly contains a set of tiles spanned by a Fibonacci tree.

### 2.2 The harp

We are now in the position to describe the main tool for the proof of theorem 1.

We use the angular sector determining a Fibonacci tree to construct a frame in which we insert the space-time diagram of a Turing machine working on a semi-infinite tape and starting its computation from an empty tape.

Such an angular sector is represented on the left-hand side of figure 3 by the two thick yellow rays supported by mid-point lines.

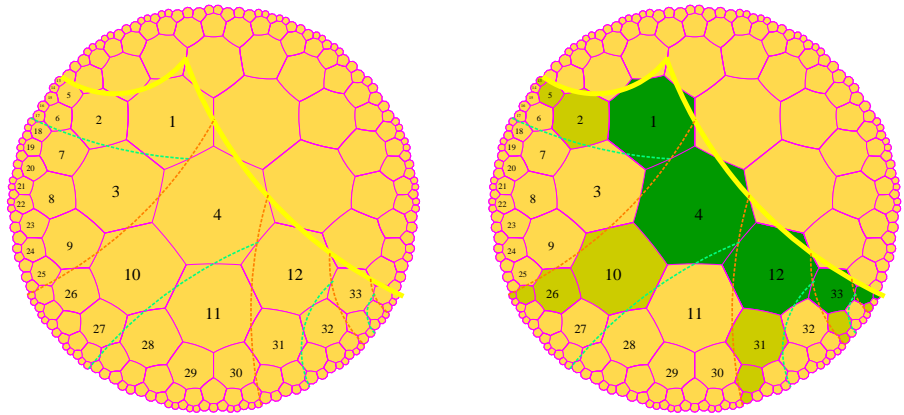

**Figure 3** *The guidelines for the harp.*

On the right-hand side of the figure, we can see the **harp** itself. It contains a Fibonacci tree which can be viewed as a space time diagram of the Turing computation. The rightmost branch of the tree, the dark green tiles on the right-hand side of figure 3, call it the **frame**, represents the Turing tape at the initial time. Define **levels** in the Fibonacci tree as the set of tiles which are at the same distance, in tiles, of the root of the tree. The distance from $T$ to the root is the smallest length of a sequence of tiles joining $T$ to the root with the condition that two consecutive terms of the sequence share a common side. Call **borders** of the tree the two mid-point lines which determine it. Note that each level meets the right-hand side border in two points, say $A$ and $B$, with, for instance, $A$ being the closest to the root. From $B$, we define the other mid-point line determined by $B$ and we consider the ray $r_B$ of this line which is issued from $B$ and which goes inside the tree. We call **chord** the set of tiles which belong to the tree, which meet $r_B$ and which are in the half-plane defined by $r_B$ which does not contain the root of the tree. A chord represents the evolution, in time, of the square of the tape: the one which is associated with the tile of the chord which is in contact with the right-hand side border of the tree.

Now, the simulation of the computation of a Turing machine goes as follows.

The starting point of the computation is the tile which bears the root of the tree: it is the single tile which is in contact with both borders.

Next, the computing signal, which bears the current state of the machine head, goes to the middle son of the root and there, on the level 1, it goes to the right as the head of the Turing machine never goes to the left of its initial position. The tile which is immediately reached on the same level is a tile of the border. At this point, it performs the required instruction.

Now, assume that the computing signal just performed an instruction: it is on a chord. The tile sends the new content of the tape to the next tile on the chord. The computing signal has now the new current state. It goes down along the chord by one level and, on the new level, it goes in the direction indicated by the instruction it has just performed. It remains on the same level until it meets the next chord, in the direction which it follows. When the chord is met, it performs the instruction. And the process is repeated.

It is not difficult that this gives an alternative solution to the origin-constrained problem, as indicated in [9]. Also in [9], we have indicated the tiles which allow to construct the harp and its simulation process.

## 3   A solution to the finite tiling problem

Now, we come to the proof of theorem 1.

We first notice that there is just a slight modification to perform on the tiles given in [9] in order to prove the theorem.

Indeed, we give a special colour to the border, call it the **silver signal**. This colour is also generated by the halting instruction if it occurs. A silver signal spreads on both sides along the level where the halting instruction was triggered. Accordingly, it meets the silver signal of the borders. At the meeting point, the signals merge in the shape of a corner. Call also **border** this part of the level delimited by the two borders of the tree. It is clear that any tile which contains the silver signal has an inward side and an outward one. We decide that the edges which delimit the outside and which are not crossed by the silver signal are blank. Accordingly, such a side may abut with the blank tile and so, this delimits the computation and outside this area, the solution which consists in tiling the area with the blank gives a solution.

And so, we proved that if the Turing machine halts, there is a tiling realized by copies of $T$ which contains at least the copy one tile of $T$ and which only contains finitely many copies of tiles of $T$.

Now, assume that there is such a tiling $\tau$, that $\tau$ contains finitely many tiles from $T$ and at least one of them. Necessarily, in $\tau$, there are tiles with a silver signal. Indeed, the tiles of $T$ which do not belong to the border have no blank edge. This property can be checked from the tiles for the harp given in [9] and the additional tiles which are given below, in figure 4. In fact, the last two rows of tiles in figure 4 are actually new tiles, appended to the set of tiles belonging to $T$. The other tiles of figure 4 replace the corresponding tiles of [9] which define the root of the tree and its both borders. Note that the new tiles of figure 4 are the tile $i$ which transform the silver signal raised by the halting state into two signals for the basis of the finite figure. One signal goes to the left and the other to the right. They meet the silver signals on the border of the tree thanks to the corner tiles which are provided by the tile $m$ for the left-hand side corner and the tile $n$ for the right-hand side one.

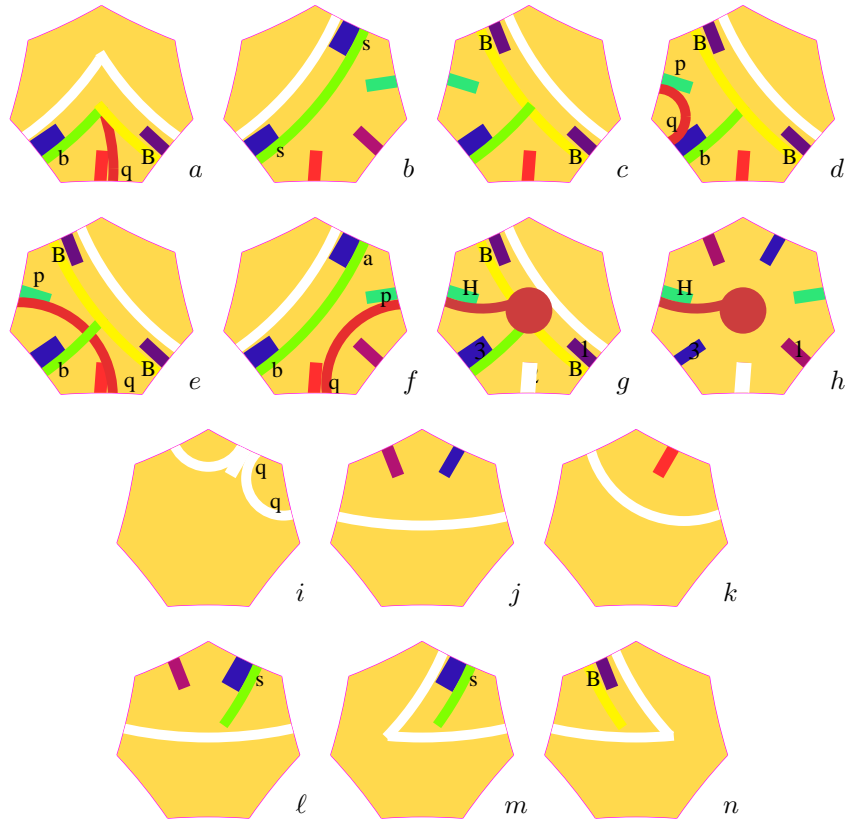

**Figure 4** *The tiles for the silver signal.*

Now, it is not difficult to see that as the non blank area of the tiling is finite, it must contain borders of the three possible kinds: left-hand side border, right-hand side one and the border on a level. If at least one component is missing, this means we have an infinite part with tiles of $T$ only. Now, as there is a left-hand side border and a right-hand side one, they must meet. And so, the tiling necessarily contains an origin. Accordingly, the tiling simulates the commputation of a Turing machine. Now, as it is finite, the simulated machine halts. And so, we proved that there is a finite solution containing at least one non blank tile if and only if the simulated Turing machine halts. And so, the problem is undecidable.

## 4 Conclusion

The finite tiling problem is used in [4] to prove that it is undecidable to know whether a cellular automaton in the Euclidean plane with Moore neighbourhood is surjective. At the present moment, there are two difficulties to transport this proof to the hyperbolic plane. The first one is the use of the classical characterization of surjective cellular automata with cellular automata whose global function is injective when restricted to finite configurations, see [14,15]. Although Hedlund's theorem can be transported to the hyperbolic plane at the price of an additional condition, see [12], the arguments of Moore's and Myhill's proofs cannot be transported to the hyperbolic plane. The second obstruction lies in the plane-filling property: it is still open whether this holds or not for the hyperbolic plane.

Accordingly, there is still much work to do in these directions.

Another point is the following. In the Euclidean plane, it is not difficult to see that, from the proof of the finite tiling problem, we get the proof thqat periodic tiling problem is also undecidable. Recall that this latter problem consists in asking whether, from a given finite set of prototiles, it is possible or not to tile the plane in a periodic way. This problem was proved undecidable by Yu. Gurevich and I. Koriakov, see, [3]. The present construction cannot be used to prove the undecidability of the periodic tiling problem in the hyperbolic plane. However, the construction of [10,13,11], combined with the construction of the present proof, gives a solution which is still not immediate. This will be the purpose of a forthcoming paper, where we also discuss the adaptation of the notion of periodicity in the hyperbolic plane, a question which is also not straightforward.